\documentclass[onecolumn,superscriptaddress,aps,pra,showpacs,floatfix]{revtex4}

\usepackage{latexsym}
\usepackage{amsmath}
\usepackage{amssymb}
\usepackage{latexsym}
\usepackage{color}
\usepackage{graphicx}
\usepackage{cancel}
\usepackage{bbm}
\usepackage{bm}
\usepackage{bbm}
\usepackage{maybemath}
\usepackage{hyperref}
\usepackage{verbatim}

\begin{document}

\title{New upper bound of muon neutrino mass in a short-baseline experiment}

\author{A. M. Attia}

\affiliation{Institute of Physics, University of Debrecen, Debrecen, Hungary}

\affiliation{Physics Department, Faculty of Science, Zagazig University, Zagazig, Egypt}

\author{I. G. M\'ari\'an} 

\affiliation{Institute for Nuclear Research, Atomki, Debrecen, Hungary}

\affiliation{Institute of Physics, University of Debrecen, Debrecen, Hungary}

\author{B. {U}jv\'{a}ri}

\affiliation{Institute for Nuclear Research, Atomki, Debrecen, Hungary}

\affiliation{Faculty of Informatics, University of Debrecen, Debrecen, Hungary}

\begin{abstract}

In the paper Int.J.Mod.Phys.E 23 (2014) 1450004, the potential of short-baseline experiments was proposed to measure the mass (and parameters of Lorentz-violating effects) of the muon neutrino, where a roughly estimated upper bound of 420 eV was given as a possibility with large unknown uncertainties. In the present work, we improve upon this study by focusing on a feasible and improved experimental setup with today's technology, eliminating most large uncertainties, with the use of the Geant4 simulation toolkit. High-energy protons collide with a tungsten target, producing a variety of particles, most importantly pions that decay into muon neutrinos. The detector records the time of flight for both muon and anti-muon neutrinos, utilizing light as a reference signal. Additionally, it captures the energy deposited by neutrinos. By applying the dispersion relation, we determine the muon and/or anti-muon neutrino mass. Our improved results reveal a less optimistic but more accurate and realistic estimated upper bound of the muon neutrino mass, providing a new limit of about 150 keV. Notably, this finding is a factor of three lower than the best upper bound previously established in the literature originating from pion decay in flight.

\end{abstract}


\maketitle

\section{Introduction}

As a result of the discovery of neutrino flavor oscillations, neutrinos are thought to have a non-zero mass, as opposed to the standard model (SM). The detection of neutrino oscillations in the atmospheric Super-Kamiokande\cite{Fukuda1999} and solar Sudbury Neutrino Observatory (SNO)\cite{Ahmad2002} experiments provided initial evidence supporting the existence of nonzero neutrino masses. In the 21st century, several neutrino oscillation experiments were conducted, providing precise measurements for the phenomenon of neutrino oscillation
\cite{Abe2008,Adamson2008,Abe2014,Choi2016,An2017,Abe2018,Abe2017,Acero2018}. 
However, these experiments revealed that two out of the three flavors are heavy, and the massive flavor possesses a mass of at least 0.05 eV. Nonetheless, it should be noted that these experiments could only determine mass-squared differences between the flavors and were unable to directly measure the individual mass of each flavor\cite{StrumiaVissani2010}. It is crucial to highlight that the theoretical concept of neutrino oscillation was initially proposed by the Russian scientist Bruno Pontecorvo to elucidate the absence of detected atmospheric and solar neutrinos\cite{Landau1957,LeeYang1957,Salam1957}. 

Measuring neutrino mass is of a great importance due to its implications regarding not only refining our understanding about the nature of the universe and dark matter, but also conceivably providing insights into some new physics beyond SM. Therefore, one of the main objectives of particle physicists has been to measure neutrino masses for several years. 
Consequently, numerous experiments have been conducted since 1991 to measure the mass of neutrinos based on tritium beta decay, from Los Alamos to the Karlsruhe Tritium Neutrino experiment in 2022. At Los Alamos, researchers established an upper limit of 11 eV at a 95\% confidence level for the mass of the electron anti-neutrino, $m(\Bar{\nu}_e)$ \cite{Robertson1991}. In the first run in 2019, the KATRIN experiment significantly improved the sensitivity of $m(\Bar{\nu}_e)$, setting a new upper bound of 1.1 eV at 90\% confidence level (CL), which represents an improvement by a factor of about two compared to the previous limit \cite{Aker2019, Aker2021}. Furthermore, in the second run in 2022, they achieved a more precise upper bound of 0.9 eV at 90\% CL. The results from the KATRIN 2019 (first run) were then combined with those from KATRIN 2022 (second run), resulting in a more accurate upper limit of 0.8 eV at a 90\% CL for $m(\Bar{\nu}_e)$ \cite{Aker2022}. 
On the other hand, the most successful attempts to measure the mass of muon neutrinos, as the second flavor, were in 1982, involving the measurement of muon neutrinos from pion decay in flight and achieving an upper bound of less than 500 keV \cite{Anderhub1982}, and in 1996, measuring muon neutrinos from the decay of pions at rest, resulting in an upper limit of less than 170-190 keV 
\cite{Assamagan1996, Bolanos-Carrera2024, NeutrinoMass2016}. 

In this paper, a short-baseline neutrino experiment is investigated for the sake of measuring muon and/or anti-muon neutrino mass $m(\nu_\mu)$ (since it is difficult to distinguish between them) using Monte-Carlo technique based on Geant4 toolkit (see Ref. \cite{GEANT4,GEANT4doc,Agostinelli2003,Amako2000,GuatelliaEtAl} for Geant4). This strategy relies on transforming time and energy deposited information of neutrinos $\nu_\mu$ into mass measurements by using the following dispersion relation, 

\begin{equation}
\label{dispersion}
    E(\nu_\mu)=\frac{m(\nu_\mu)c^2}{\sqrt{1-v^2/c^2}}
\end{equation}
where $v$ and $c$ are the speed of neutrinos $\nu_\mu$ and light, respectively, while $m(\nu_\mu)$ and $E(\nu_\mu)$ are the mass and deposited energy of neutrinos $\nu_\mu$ reaching the detector. In other words, once we get the time of flight (TOF) of both $\nu_\mu$ and light, as well as $E(\nu_\mu)$ hitting the detector, the dispersion relation in question yields $m(\nu_\mu)$.

A short-baseline experiment has many advantages. For one, it makes the distance measurements much easier compared to long baseline experiments. 
Another key advantage is the possibility to construct a tiny vacuum pipe for a laser pulse in the middle of the baseline. This pulse can be used as a reference for more precise time measurements comparing directly the time of flight and consequently the speed of photons and neutrinos.
Such relatively short vacuum tunnels for laser pulses are already in use and operation mainly in interferometers, e.g. in the LIGO, Virgo, KAGRA experiments \cite{LIGO2015,Virgo2015,Kagra2021}. Therefore a decommissioned interferometer would be a great opportunity to place our proposed neutrino experiment reusing the already built vacuum pipes.

This time of flight experimental setup can also be used to constrain the Lorentz-violating parameters with a direct measurement and test for the Lorentz-violation of neutrinos in a labortory environment. For more detailed descriptions of Lorentz-violation see \cite{Jentschura2019,Somogyi2019,Jentschura2020}, however in this work we focus on measuring the mass of the muon neutrino, since threshold considerations give very strong constraints for the Lorentz-violating parameter $\delta_\nu$, see Eq.~(31) of \cite{Jentschura2020}.

In \cite{Jentschura2014} these basic concepts of the experiment were proposed with a review of the theoretical status. Here the center of attention is on the feasibility and practicality of the experiment with today's technology. 
An improved setup is presented with the following main additional features.
(i) The fact that the created neutrinos have a wide energy range is taken into account. For this issue we propose a solution by considering the track lengths of the created muons in the crystal scintillator detector.
(ii) The threshold energy of neutrino detection is also included. 
(iii) The finite size of the vacuum chamber is considered. Particles hitting the walls of the vacuum chamber can also create neutrinos that can reach the detectors. We show that with an appropriate time cut the contribution of these neutrinos are negligible.
(iv) Instead of a point-like source, we have a more realistic beam length. Luckily with clever beam monitoring tools, using Cherenkov detectors, this is not a significant limiting factor.

Finally let us not that, as it was the case for electron neutrinos, in the future, better equipment could significantly improve the results described in this work.

The paper is structured as follows: Following the introduction, the second section details the experimental setup. The third section examines the optimization of key parameters essential to the experiment, such as target width and magnetic field strength. Subsequently, the fourth section discusses time of flight (TOF) calibration, while the fifth section investigates the interaction rate of neutrinos at the crystal scintillator detector. Moving forward, the sixth section addresses the detection of neutrinos with an energy cut. The seventh section involves calculating the mass of muon neutrinos. Finally, the paper concludes by summarizing the findings and suggesting potential enhancements based on forthcoming technological advancements.

\section{Experimental setup}\label{Experiment setup}
The experiment must be designed to produce neutrinos with high efficiency. 

For this reason high energy protons are injected into a relatively short ($\sim$100 m) vacuum chamber with a kicker.
In the same time the proton kicker triggers the reference laser pulse. The protons then hit the target producing many particles, where the most relevant ones in this experiment are the pions that quickly decay to muon neutrinos, e.g. $\pi^+ \to \mu^+ + \nu_\mu $. In order to eliminate the unwanted charged particles and the pions that decay with a significant time delay, a strong magnet is placed in the vacuum chamber. Then the race begins between the neutrinos and the well timed laser pulse. The timing is helped and corrected by Cherenkov detectors placed at the end of the vacuum chamber. The final destination of the neutrinos is a crystal scintillator detector and a photodioode for the laser pulse, measuring the difference between their TOF.

\begin{figure}[h]
    \centering
    \includegraphics[width=1.0\linewidth, height=0.25\textheight]{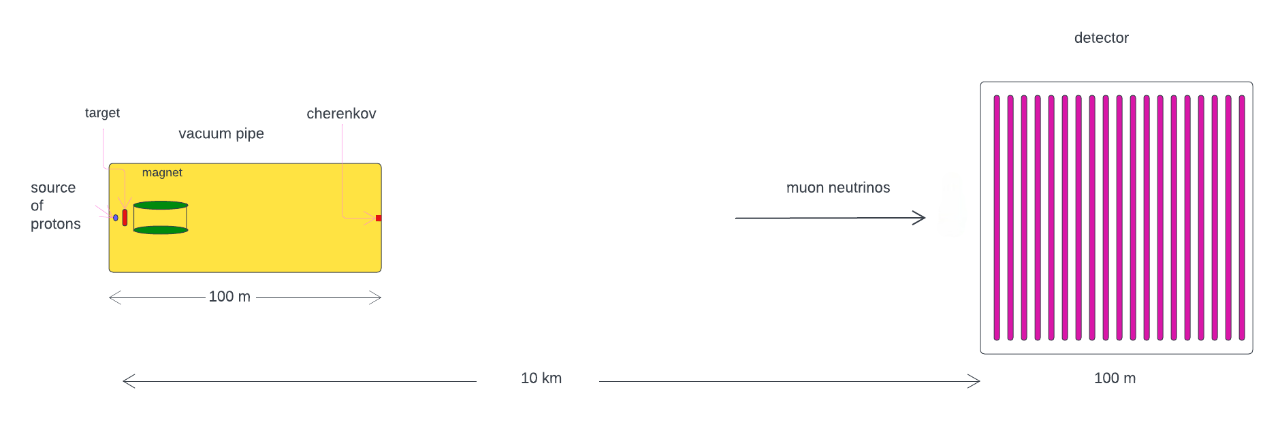}
    \caption{Sketch of the proposed experiment}
    \label{fig:experiment}
\end{figure}

Since we adopted a short-baseline experiment, its dimensions have been designed to be 10 km long. The simulation of the above described setup must contain a particle gun, target, magnet, rock and crystal scintillator detector as the five major elements of this experiment as shown in Fig.~\ref{fig:experiment}. 
Since we are interested in producing neutrinos, one approach is to collide protons with a target. Therefore, the particle gun is constructed, positioned 1 meter away from the target in the -z direction, to generate protons.

The protons are provided by a high energy proton accelerator such as the Super Proton Synchrotron (SPS) which possess the following properties: a bunch intensity in the order of $1.1 \times 10^{11}$ protons per bunch, an energy of 450 GeV, and a bunch length of 500 ps\cite{Tahir2008,Lyndon2007,Tahir2007}.

The target is made of tungsten due to the following reasons: tungsten has high atomic number, i.e.,  high number of protons are contained in its nucleus and as a result various particles are produced such as pions and kaons which undergo a subsequent decay to neutrinos. Additionally, tungsten's density allows it to withstand not only highly energetic collisions, but also the associated heating, making it a good choice for the target material\cite{Grimvall1987,Nogami2021}. It has a width of 10 cm in the z-direction, while its dimensions along the x and y axes are 40 $\times$ 40 cm.  The target is positioned with its center located at 5 cm along the z-direction, i.e., its front and back sides along z direction lies at the origin of coordinates and 10 cm, respectively.

Following the target is a magnet with a strength of 4 T and is designed as a cylinder with a height of 1 m and a radius of 1.5 m. 
The magnetic field is utilized to deflect low-energy, i.e., slow outgoing particles in either the positive or negative x-direction away from the detector, particularly those capable of decaying into neutrinos with a significant delay in their TOF. The neutrinos produced by those low energy particles will arrive late at the detector, and yield not precise TOF measurements.

The proton gun, target, and magnet are placed inside a vacuum pipe to to ensure the prevention of particles' collisions. Possibly positioned underground, this pipe is surrounded by earth, which is simulated using SiO$_2$ (silicon dioxide) rock. The vacuum pipe's measurements extend to 40 m in the x-direction, 40 m in the y-direction, and 100 m in the z-direction.

Ultimately, a crystal scintillator detector with dimensions of 100 meters along the three coordinates is placed in such a way that its front is located at 10 km along the z-direction, which corresponds to coordinates (0, 0, 10) km, at the end of the setup. The size of the detector is chosen to ensure that there are sufficient statistics for presentation. However, in practice, dimensions of 20 $\times$ 20 $\times$ 100 meters along the x, y, and z axes, respectively, are preferable in order to reduce costs.

\section{Optimization of key parameters}

The beam energy, bunch length, target width, and magnetic field are crucial parameters in this experiment. To enhance our results and reduce uncertainties, we refined and optimized these parameters as follows.

Firstly, a thicker (10 cm) target width in z direction is chosen which yields large number of pions, which subsequently decay into neutrinos. The reason for aiming at high number of neutrinos is that neutrinos are weakly interacting and have very low cross section, so choosing a thicker target will enable us to increase our chance for detection.
We found that this thickness still does not compromise the TOF measurement, since high energy  protons travel close to the speed of light, therefore they do not add a delay in time while they travel trough the target. If scattering occurs, then the scattered protons or pions change direction therefore neutrinos created by them rarely reach the detector.

In the present simulation we fixed the beam energy and bunch length to the current capabilities of the SPS facility. 
Our aspiration for the future is the construction of a high-energy proton beam accelerator with an approximate zero bunch length. This will be perhaps achievable using plasma wakefield accelerators utilizing ultra-short laser pulses. 

\begin{figure}[h]
    \centering
    \includegraphics[width=0.8\linewidth, height=0.37\textheight]{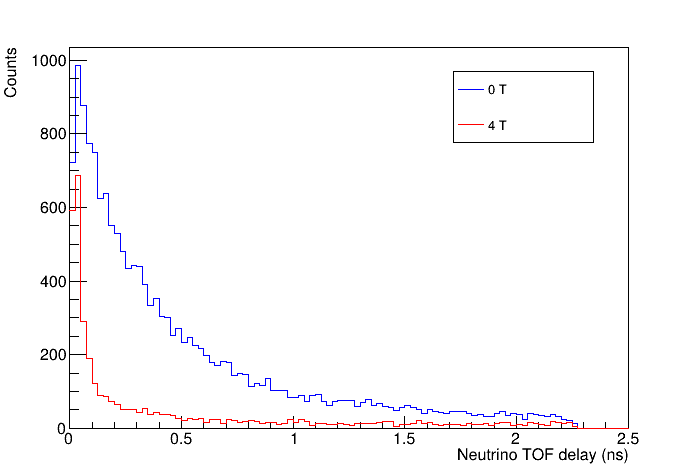}
    \caption{Neutrino TOF delay in the case of 4 T (red) and zero magnetic field (blue) as compared to the time it takes for light in vacuum to reach the crystal scintillator, with energy $<$ 1 GeV.}  
    \label{fig:time_0T_4T}
\end{figure}

A magnetic field strength of 4T is specifically employed to deflect low-energy particles that might otherwise decay into neutrinos at a later stage. These late neutrinos are not of the focus of interest in order to ensure better timing resolution and hence, a precise mass measurement. The delay in the TOF compared to the TOF of the laser pulse, both with and without the use of a magnetic field, is examined for neutrinos with energies less than 1 GeV as illustrated in Fig.~\ref{fig:time_0T_4T}.  As depicted from the Fig.~\ref{fig:time_0T_4T}, the TOF distribution of neutrinos in the presence of a magnetic field (B = 4T) is indeed notably narrower than that observed with a magnetic field strength of 0T. However the peak also becomes smaller for larger magnetic fields resulting a smaller number of relevant events. On the other hand the prize of magnets drastically increase with their strength above 4T, therefore we choose 4T as the optimum value.

It is worth noting that Fig.~\ref{fig:time_0T_4T} highlights the advantages of using the magnet. However, we will later employ a more precise TOF cut to select the first neutrinos that hit the detector, primarily originating either from interactions of pions in the target or pions' decay in flight, rather than from the rock. 

\section{Timing calibration} 
\label{Timing calibration}

Since the measurement of neutrino mass relies on time of flight (TOF) measurements, performing precise timing calibration is a critical task to reduce uncertainty. In practice, the proton beam striking the target has a bunch length of 500 picoseconds, meaning that not all protons arrive at the target simultaneously. Furthermore, the arrival times of both the front and back of the proton bunch are not well-defined.

\begin{figure}[h]
    \centering
    \includegraphics[width=0.8\linewidth, height=0.37\textheight]{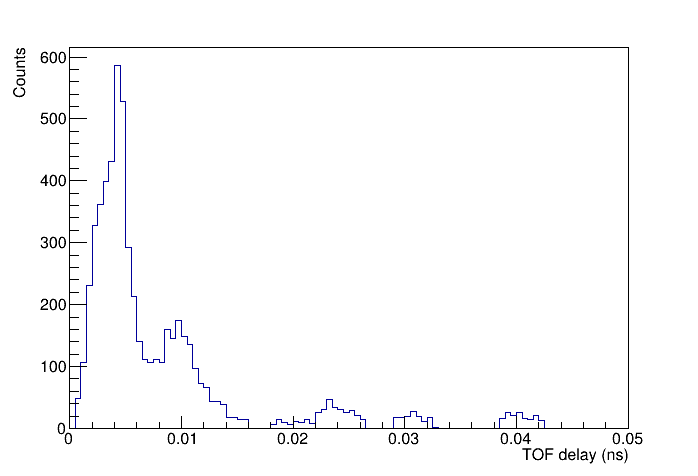}
    \caption{Number of photons produced as a result of charged particles hitting the Cherenkov detector in case of a point-like proton source.}
    \label{fig:cherenkov}
\end{figure}

In our timing calibration process, we employed a two-step approach. First, we calibrated the time uncertainty of the incident proton beam. To achieve this, we placed a small Cherenkov detector measuring 1 cm x 1 cm x 1 cm at the end of the vacuum pipe, situated approximately 100 meters away from the target along the z-axis and oriented at a 0° angle relative to the z-axis.
Fig.~\ref{fig:cherenkov} shows the signal of the Cherenkov detector in case of a 0ps bunch length for $14.75 \times 10^6$ protons. 
In a real experiment the number of detected pions would be orders of magnitude larger resulting in a much better statistics which would allow for more sophisticated analysis methods, however in this simulation generating the accurate number of events requires an unreasonable computational time.
The signal has a well defined first peak with a full width at half maximum of around 5ps. In cases when the bunch length differs from zero, the produced signal is the convolution of the bunch density in the z direction and the signal of the Cherenkov depicted in Fig.~\ref{fig:cherenkov}. 
For instance, the width of the TOF measurements for bunch lengths of 20, 100, and 500 picoseconds is around 25, 105, 505 ps which includes the additional 5ps uncertainty of the zero bunch length signal. Consequently, the signal of neutrinos detected 10 km away by the crystal scintillator is also a convolution of the bunch length and the zero bunch length signal.  Therefore, applying a deconvolution to the crystal detector's signal using the Cherenkov signal would yield an approximation of the same function as we obtain in the case of a zero bunch length resulting only a 5ps uncertainty even for large bunch lengths. For this reason we can restrict our simulations using only a point source adding the 5ps uncertainty at the end of the analysis. This approach is also driven by the hope that future accelerator projects will be able to provide us with a high-energy proton beam source that closely resembles a point source.

\begin{figure}[h]
    \centering
    \includegraphics[width=0.8\linewidth, height=0.37\textheight]{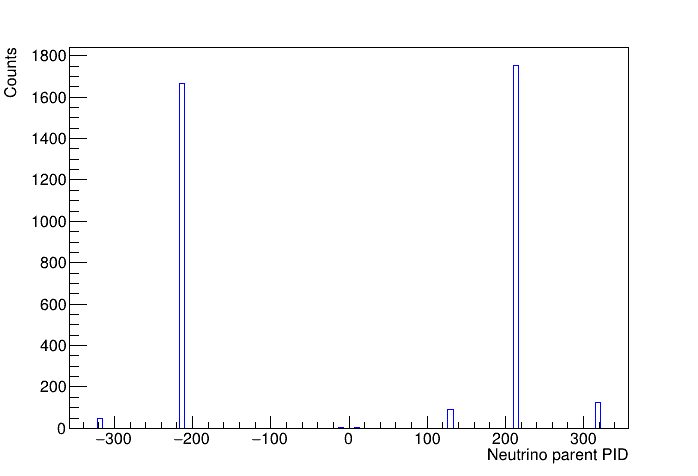}
    \caption{PID of neutrinos' parent, where the PID values -321, -211, 13, 130, 211, and 321 correspond to $K^-$, $\pi^-$, $\mu^+$, $K^0_L$, $\pi^+$, and $K^+$, respectively.} 
    \label{fig:parentID}
\end{figure}

\begin{figure}[h]
    \centering
    \includegraphics[width=0.8\linewidth, height=0.37\textheight]{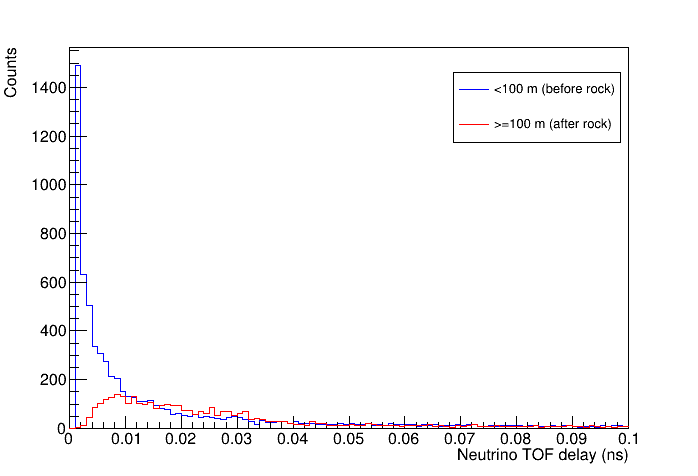}
    \caption{TOF delay for neutrinos generated before (blue) and inside the rock (red) upon reaching the crsytall scintillator detector, with $dt< 3.4$ ns} 
    \label{fig:time_before_after_rock}
\end{figure}

Next, we addressed the time delay associated with pions, the primary source of neutrinos as depicted in Fig.~\ref{fig:parentID}, which shows that the parent of the majority of neutrinos are pions in the target due to multiple scatterings. This involved examining the time uncertainty in the arrival distribution of neutrinos at the scintillator detector. Since the fastest neutrinos are either produced in the target or originate from the decay of pions in flight within the vacuum pipe before the rock, it was crucial to establish that the majority of neutrinos arriving before the rock indeed outweighed those arriving slightly later from the rock. Figure \ref{fig:time_before_after_rock} demonstrates that the majority of the fastest neutrinos reaching the scintillator detector are produced before the rock. Additionally, the time delay for these fastest neutrinos, depicted by the peak in blue, is approximately 5 ps.

In conclusion, the total uncertainty, dt, is the sum of the calibrations for the proton beam and the neutrino beam, amounting to 10 ps. This calibration was executed with the assistance of both the small Cherenkov detector and the scintillator detector.

\section{Neutrino interaction rate in the detector}

In order to see whether this experiment with all of the applied cuts produce enough events, let us examine the interaction rate in the detector, i.e. how much time one has to wait to collect enough useful events.

As mentioned earlier in Section \ref{Experiment setup}, the most optimal location for the crsytall scintillator detector is along the z-axis, precisely at coordinates (0, 0, 10) km from the origin. This choice is the best one, since this ensures that the neutrinos detected are the ones that were produced either close to or within the target, without any additional delay caused by the time it takes for their parent particles to decay.

\begin{figure}[h]
    \centering
    \includegraphics[width=0.8\linewidth, height=0.37\textheight]{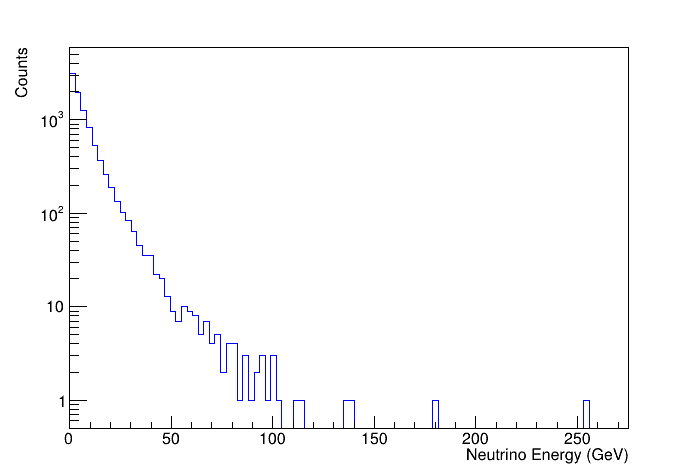}
    \caption{Energy distribution (in log scale) of neutrinos arrived at the crystal scintillator detector with dt $<$ 3.4 ns.}
    \label{fig:energy}
\end{figure}

Since the interaction rate of neutrinos with the crystal scintillator detector depends on their energies, figuring out their energy distribution is a crucial step in our analysis (refer to Fig.~\ref{fig:energy}). In this plot, we implement a TOF delay cut for neutrinos relative to the TOF of light in a vacuum, with a threshold of less than 3.4 ns. As a practical adjustment for the scintillator detector, we can achieve this by ceasing the collection of hits beyond that time.

The interaction rate of neutrinos, i.e., the number of neutrinos interacting per second with detector material, is determined by the following formula\cite{Diwan2018,DiwanSpring2018}:
\begin{equation}
	R=l \, \rho_T \int dE \, n_\nu(E) \, \sigma(E)
\end{equation}
where $R$ is the event rate, $l$ is the path length, $\rho$ is the target density, $n_\nu(E) dE$ is the number of neutrinos created in the energy range $E+dE$, and $\sigma(E)$ is the cross section for neutrino energy $E$.
 
The typical cross-section is given by $10^{-38}$ cm$^2$ times the neutrino energy (in GeV). The target density consists of the particle targets within the detector, specifically the number of protons and neutrons, as we neglect electrons due to their small size. As a rough estimate, there are approximately $6\times 10^{29}$ protons and neutrons in one ton of detector material, which means $6\times 10^{23}$ per cm$^3$.

\begin{figure}[h]
    \centering
    \includegraphics[width=0.8\linewidth, height=0.37\textheight]{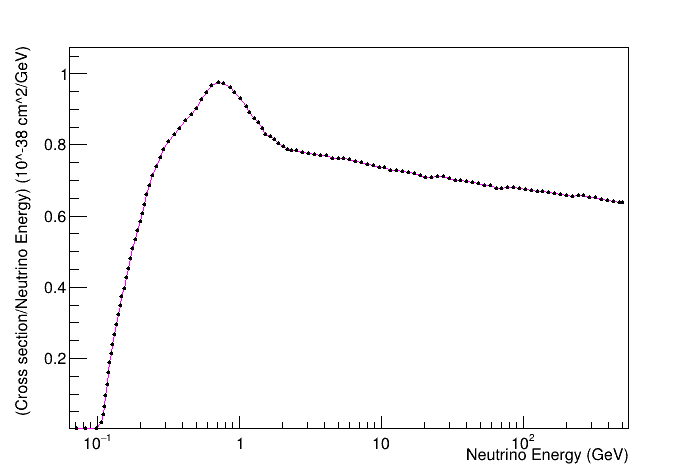}
    \caption{Total neutrino cross section per neutrino energy as a function of neutrino energy \cite{Formaggio2012}.}
    \label{fig:crossSection}
\end{figure}

In our analysis, we calculated the interaction rate of neutrinos per proton as it is described by the following formula,

\begin{equation}
\label{interactionRatePerProton}
	R_p=\frac{l \, \rho_T}{N_p} \sum_i \, N_\nu(E_i) \, \frac{\sigma(E_i)}{E_i} E_i
\end{equation}
where $R_p$ is the event rate per proton, $N_p$ is the number of protons in the simulation, $l$ is the path length, $\rho_T$ is the target density, $N_\nu(E_i)$ is the number of neutrinos created in the energy bin $E_i$, $\sigma(E_i)/E_i$ is the total neutrino cross section per neutrino energy, which can be read off from Fig.~\ref{fig:crossSection} in $10^{-38}$cm$^2$/GeV, $E_i$ is the i-th energy bin, and the summation goes over all energy bins. 
The number of neutrinos in each energy bin is obtained from the results of our simulation depicted in Fig.~\ref{fig:energy}, the path length is determined by the detector's geometry and is set to 10,000 cm, the target density represents the number of protons and neutrons, approximately $6\times 10^{23}$ per cm$^3$, and the no of protons refers to the number of protons used in this simulation, which is $14.75 \times 10^6$ protons. In this simulation, we focus on low energy neutrinos with an energy of 150 MeV and according to Eq.~\ref{interactionRatePerProton}, the interaction rate per proton for such neutrinos is approximately 4.58336 $\times 10^{-16}$, assuming an ideal detector response.

\begin{figure}[h]
    \centering
    \includegraphics[width=0.8\linewidth, height=0.37\textheight]{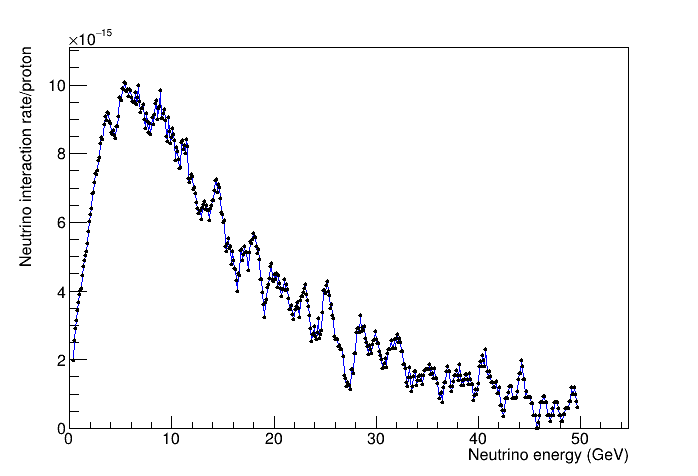}
    \caption{Neutrino interaction rate per proton for different neutrino energies.
     }
    \label{fig:interaction_rate}
\end{figure}

Since the Super Proton Synchrotron produces approximately 1.1 $\times 10^{11}$ protons per bunch and operates at a bunch frequency of about 1 MHz, the expected number of neutrinos with an energy of approximately 150 MeV detected per second can be calculated as follows: Event rate per proton (4.58336 $\times 10^{-16}$) × number of protons per bunch (1.1 $\times 10^{11}$) × bunch frequency (1 $\times 10^6$), which yields roughly 50. This means that approximately 50 events per second can be obtained from the interaction of 150 MeV neutrinos with the scintillator detector.\\
\\

\section{Detection of neutrinos with an energy cut}

It is important to focus on neutrinos that have a relatively low energy, since from Eq.~\ref{dispersion} it is clear that the achievable mass threshold is directly proportional to the energy of the measures neutrino. Therefore one must find a method to select for low energy ($<150$ MeV) neutrinos while discarding the high energy ones.

Muon neutrinos contribute to various types of interactions depending on their energy. In the case of low energies, specifically those less than 2 GeV, Quasi-Elastic Scattering dominates \cite{Formaggio2012,Ba2023priv}. This process converts a proton of the detector material into a neutron with the emission of a $\mathrm{\mu^-}$ (muon) in the final state for muon neutrinos, and it converts a neutron into a proton, accompanied by a $\mathrm{\mu^+}$ (muon), for muon antineutrinos.

\begin{figure}[h]
    \centering
    \includegraphics[width=0.8\linewidth, height=0.37\textheight]{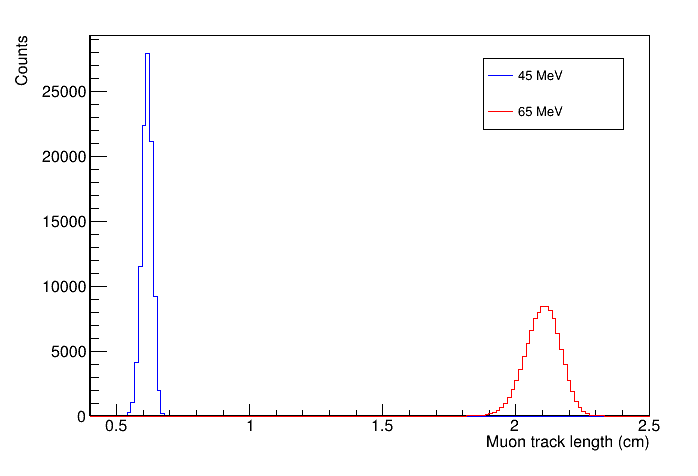}
    \caption{Distance covered by a 45 MeV (blue) and 65 MeV (red) muon until it reached its stopping point within the detector material.}
    \label{fig:muonTrack}
\end{figure}

Given our focus on neutrinos with an energy of 150 MeV, the muons produced as a result of this interaction typically possess a kinetic energy of around 45 MeV and travel a distance of approximately 0.6 cm from their creation point before coming to a stop in the detector material (as illustrated in Fig.~\ref{fig:muonTrack}). If the path length exceeds this threshold, it is likely that the muon produced originated from neutrinos with energies greater than 150 MeV, and consequently, such events can be excluded. As shown in Fig.~\ref{fig:muonTrack} for a muon energy that is 65 MeV the stopping distance is much greater, without any overlap between the peaks which allows a precise energy cut. Small crystal scintillator detectors, already in use, utilize LYSO (Lutetium-yttrium oxyorthosilicate) scintillators, commonly employed for detecting high-energy particles owing to their excellent light output properties, with dimensions typically measured in a few millimeters \cite{{CMS-TDR-020}}. Therefore one can set the trigger for a useful event to be only a few nearby detectors going off in a cubic detector array. A detector array can also eliminate cosmic muons by discarding signals that trigger the detectors on the surface of the array. Another important use of this type of detector is the possibility to calculate the elapsed time the neutrino took in the detector array before it interacted with the first detector. Therefore the time resolution of the detector, i.e. the time uncertainty will be the size of a small detector divided by the speed of neutrinos, which is in the order of a few picoseconds. 

The second type of neutrino interaction is called Resonant Single Pion Production or Coherent Pion Production, where a pion is produced in the final state along with the muon. This interaction dominates in the energy range above 2 GeV and less than 20 GeV \cite{Formaggio2012,Ba2023priv}. However, there is still a lesser chance for neutrinos with energies between 2 GeV and 20 GeV to produce muons with an energy of 150 MeV. These 150 MeV muons can be identified by the detector and then disregarded, as in this case, these muons are accompanied by the production of a pion in the final state.

On the other hand, if the energy exceeds 50 GeV, Deep Inelastic Scattering (DIS) plays a significant role \cite{Formaggio2012,Ba2023priv}. The number of neutrinos with energies in this range is pretty low as depicted in Fig.~\ref{fig:energy}, and there is also a reduced likelihood for these neutrinos to produce a muon with an energy of 150 MeV. Furthermore, in DIS, the final state typically consists of a muon and a shower of hadrons, which can be easily identified by the detector. Therefore, we can confidently identify and disregard the muon produced with an energy of 150 MeV as originating from DIS.

\section{Muon neutrino mass calculation}

Considering that the time resolution of the crystal scintillator detector is approximately 8 ps, and the time uncertainty of the proton beam hitting the target is about 5 ps, as well as an additional 3 ps due to the time delay of pions in the target. So, the total time uncertainty, denoted as $\delta t$, amounts to 16 ps. Consequently, the ratio of the deviation in the TOF of neutrinos compared to the TOF of light ($t_0$) can be expressed as follows:
\begin{equation*}
    |\delta| \approx \frac{\delta t}{t_0} = c \frac{\delta t}{s_0}=\frac{(3 \times 10^8)(16 \times 10^{-12})}{10000}\approx 5 \times 10^{-7}
\end{equation*}
where s$_0$ is the length of the baseline which is 10 km. Then the mass of muon neutrino from Eq.~\ref{dispersion} to the first order, i.e., neglecting the $O(\sqrt{\delta^2})$ terms is given by
\begin{equation*}
    m(\nu) \approx \frac{E_{\nu}}{c^2} \sqrt{2|\delta|} \approx (150 \times 10^6)\sqrt{2 \times 5 \times 10^{-7}}= 150 \, \mathrm{keV}
\end{equation*}
The obtained result is three times lower than the value reported by researchers in 1982 for neutrinos originating from the decay of pions in flight, where they established an upper limit of 500 keV for the muon neutrino mass \cite{Anderhub1982} 

\section{Conclusions}
\label{sec4}
We designed and simulated a short-baseline neutrino experiment with the focus of measuring the mass of muon and/or anti-muon neutrinos. Our approach involves utilizing a novel method to determine the mass of muon and/or anti-muon neutrinos based on their time of flight, employing the dispersion relation measuring the speed of neutrinos compared to a laser pulse. The majority of muon and/or anti-muon neutrinos in our experiment are generated through the decay of pions in flight.

The most notable achievement in measuring the mass of muon neutrinos from the decay of pions in flight occurred in 1982, with a value of 500 keV. However, through our new method, simulation, and calculations, we have successfully established an upper bound of approximately 150 keV. This upper bound represents a significant improvement, being three times lower than the previously mentioned best upper bound.

Furthermore, our experiment brings muon neutrinos back into focus, as the scientific community has predominantly focused on electron neutrinos in recent times. Our work not only achieves an enhanced upper bound, but also contributes to cost reduction due to the short-baseline nature of the experiment.

A large part of the cost comes from creating a tiny vaccuum tunnel through the baseline of the experiment from the target to the final detector for the laser pulse.
Such relatively short vacuum tunnels for laser pulses are already in use and operation mainly in interferometers, e.g. in the LIGO, Virgo, KAGRA experiments. Therefore a decommissioned interferometer would be a great opportunity to place our proposed neutrino experiment reusing the already built vacuum pipes.

Naturally, as it happened in the case of measuring the mass of the electron neutrino, future technologies and advancements can massively help lowering the uncertainties of the measurements.
In our case the largest error comes from the size of the scintillator detector, therefore in the future when scintillator detectors become smaller the time uncertainty linearly decreases with it. Smaller scintillator detectors not only could reduce the time uncertainty but would also allow energy cuts smaller than 150 MeV which would significantly improve the final result.

\section*{Acknowledgments}
We would like to express our gratitude for the valuable contributions of U. D. Jentschura and I. N\'andori, with whom we engaged in insightful discussions on a weekly basis. 
Special appreciation is extended to G. Barenboim from Universitat de València-CSIC, for his invaluable private communications that provided valuable insights into potential neutrino interactions. We also wish to acknowledge and thank Milind Diwan from Brookhaven National Laboratory for his guidance and assistance during private communications related to the calculation of neutrino interaction rates.

\end{document}